\documentclass[aps,prc,twocolumn,showpacs]{revtex4}

\begin{document}

\title{Analytical expressions of the dispersive contributions of the
  nuclear optical model}

\author{J.M.~Quesada}
\email[]{quesada@us.es}
\affiliation{Departamento de F\'{\i}sica At\'omica, Molecular y
  Nuclear, Universidad de Sevilla, Ap. 1065, E-41080 Sevilla, Spain}
\author{R.~Capote}
\altaffiliation{Permanent address: Centro de Estudios Aplicados al Desarrollo Nuclear, 
Ap. 100, Miramar, La Habana, Cuba}
\affiliation{Departamento de F\'{\i}sica Aplicada, Universidad de Huelva,
 E-21071 Huelva, Spain}
\author{J.~Raynal}
\affiliation{4 rue du Bief, 91380 Chilly-Mazarin, France}
\author{A.~Molina}
\affiliation{Departamento de F\'{\i}sica At\'omica, Molecular y
  Nuclear, Universidad de Sevilla, Ap. 1065, E-41080 Sevilla, Spain}
\author{M.~Lozano}
\affiliation{Departamento de F\'{\i}sica At\'omica, Molecular y
  Nuclear, Universidad de Sevilla, Ap. 1065, E-41080 Sevilla, Spain}

\date{\today}

\begin{abstract}
Analytical solutions of dispersion relations in the nuclear optical
model have been found for both imaginary volume and surface
potentials. A standard Brown-Rho shape has been assumed for the volume
imaginary term and a Brown-Rho shape multiplied by a decreasing
exponential for the surface contribution. The analytical solutions
are valid for any even value of the exponent appearing in these
functional forms.
\end{abstract}


\maketitle

A significant contribution to the optical model theory during the last two
decades can be considered the work of Mahaux and co-workers on dispersive
optical model analysis\cite{mang84,masa87np,johoma87,masa91,masa91rev}. The
unified description of nuclear mean field in dispersive optical model is
accomplished by using a dispersion relation, which links the real and
absorptive terms of the optical model potential (OMP). The additional
constraint imposed by dispersion relation helps to reduce the ambiguities in deriving
phenomenological OMP parameters from the experimental data.\\
In a dispersion relation treatment, the real central potential strength
consists of a term which varies slowly with energy $E$, the so called
Hartree-Fock (HF) term, $V_{_{HF}}(r,E)$, plus a correction term, $\Delta V(r,E)$, 
which is calculated using a dispersion relation. The depth of the
dispersive term of the potential $\Delta V(r,E)$ can be written as 
\begin{equation}
\Delta V(r,E)=\frac {\mathcal{P}}\pi {\int_{-\infty }^\infty }
\frac{W(r,E')}{E'-E}dE'  \label{integral}
\end{equation}
Being $W(r,E)$ the imaginary part of the OMP. Assuming that $W(r,E=E_{_F})=0$ and 
$\Delta V(r,E=E_{_F})=0$, where $E_{_F}$ is the Fermi energy, then equation (\ref{integral})
can be written in the substracted form: 
\begin{equation}
\Delta V(r,E)={\frac{\mathcal{P}}\pi \int_{-\infty }^\infty {W(r,E')}
\left(\frac {1}{E'-E}-\frac {1}{E'-E_F}\right)dE'}
\label{integral_subs}
\end{equation}
With the assumption that $W(r,E)$ be symmetric respect to the Fermi energy, the
equation (\ref{integral_subs}) can be expressed in a form which is stable
under numerical treatment \cite{dewara89}, namely: 
\begin{equation}
\Delta V(r,E)=\frac 2\pi (E-E_{_F}){\int_{E_{_F}}^\infty }\frac{W(r,E')-W(r,E)}
{(E'-E_{_F})^2-(E-E_{_F})^2}dE'
\label{integral_num}
\end{equation}
The dispersive term $\Delta V(r,E)$ is divided into two terms $\Delta
V_{_V}(r,E) $ and $\Delta V_{_S}(r,E)$, which arise through dispersion relations
(\ref{integral}) from the volume $W_{_V}(r,E)$ and surface $W_{_S}(r,E)$
imaginary potentials respectively. If the imaginary potential geometry is energy 
dependent, the radial dependence of the dispersive correction can not be 
expressed using a Woods-Saxon form factor $f_{_{WS}}(r)$. However, to simplify the problem, 
the OMP geometry parameters are usually assumed to be energy independent. In
this case, $\Delta V(r,E)=\Delta V(E)f_{_{WS}}(r)$ and the energy
dependence of the real volume $V_{_V}(E)$ 
and surface $V_{_S}(E)$ parts of the dispersive OMP are given by:
\begin{eqnarray}
&&V_{_V}(E)=V_{_{HF}}(E)+\Delta V_{_V}(E)  \nonumber \\
&&V_{_S}(E)=\Delta V_{_S}(E)
\end{eqnarray}
It is useful to represent the variation of surface $W_{_S}(E)$ and volume $W_{_V}(E)$
absorption potential depths with energy in functional forms
suitable for the dispersive optical model analysis. An energy dependence for
the imaginary volume term has been suggested in studies of nuclear matter
theory \cite{brrh81}: 
\begin{equation}
W_{_V}(E)=A_{_V}~\frac{(E-E_{_F})^n}{(E-E_{_F})^n+(B_{_V})^n}
\label{volumen}
\end{equation}
where $A_{_V}$ and $B_{_V}$ are undetermined constants. Brown and Rho
\cite{brrh81}, propose $n=2 $, while Mahaux and Sartor 
\cite{masa87np} have suggested $n=4$.\\ 
An energy dependence for the imaginary-surface term has been suggested
by Delaroche \emph{et al} \cite{dewara89} to be:
\begin{equation}
W_{_S}(E)=A_{_S}~\frac{(E-E_{_F})^m}{(E-E_{_F})^m+(B_{_S})^m}~\exp
(-C_{_S}|E-E_{_F}|)  \label{surface}
\end{equation}
where $m=2,4$ and $A_{_S},B_{_S}$ and $C_{_S}$ are undetermined constants.\\
According to equations (\ref{volumen}) and (\ref{surface}) the imaginary
part of the OMP is assumed to be zero at $E=E_{_F}$ and nonzero everywhere
else. A more realistic parameterization of $W_{_V}(E)$ and $W_{_S}(E)$ forces
these terms to be zero in some region around the Fermi energy. A physically
reasonable energy for defining such a region is the average energy of the
single-particle states $E_p$ \cite{masa91}.\\
Therefore a new definition for imaginary part of the OMP\ can be written as:
\begin{equation}
W_{_V}(E)=\left\{ 
\begin{array}{lrcl}
0 & E_{_F}~< & E & <~E_{_P} \\ 
A_{_V}~\frac{{(E-E_{_P})^n}} {{(E-E_{_P})^n+(B_{_V})^n}} &  & E & \ge ~E_{_P}
\end{array}
\right.  \label{WV3}
\end{equation}
\noindent and likewise for surface absorption.
\begin{widetext}
\begin{eqnarray}
&&W_{_S}(E)=\left\{
\begin{array}{lrcl}
0 & E_{_F}~<&E&<~E_{_P} \\ 
A_{_S}~ e^{-C_{_S}|E-E_{_P}|} ~\frac{{(E-E_{_P})^m}}
{{(E-E_{_P})^m + (B_{_S})^m}} & &E&\ge ~E_{_P}
\end{array}
\right. 
\label{WS3}
\end{eqnarray}
\end{widetext}
The symmetry condition
\begin{equation}
W(2E_{_F}-E)=W(E)  \label{symmetry}
\end{equation}
is used to define imaginary part of the OMP for energies below the Fermi
energy. Equations (\ref{WV3}-\ref{symmetry}) are used to describe imaginary
absorptive potential in this contribution.\\
Nonlocality effects on the imaginary part can be treated according to
the work of Mahaux and Sartor \cite{masa91} and the analytical
solutions of their contributions coming from the dispersion relations
have been found by VanderKam {\em et al.} \cite{vaweto00}.\\

In a recent work we have presented a general numerical solution of the
dispersion integral relations between the real and the imaginary parts
of the nuclear optical potential \cite{camoqu01}. In this work we develop
analytical solutions of dispersion relations, 
equation (\ref{integral_subs}) for usually employed functional forms of the
imaginary potential, $W(E)$, used in dispersive optical model analyses
(equations (\ref{WV3}-\ref{symmetry})). According to \cite{vaweto00} we introduce the quantities $E_{_0}$ $=E_{_P}-E_{_F}$ (the offset
energy), $E_{+}=E_x+E_{_0}$ and $E_{-}=E_x-E_{_0}$.  $U$ and  $E_x$
are the integration variable and the excitation energy ($E-E_{_F}$), respectively. In the next 
formulae, we will drop the subscripts on the constants $A,B,C$.\\
For the surface potential $W_{_S}(E)$, given by equations (\ref{WS3})
and (\ref{symmetry}), we can write the dispersive integral
(\ref{integral_subs}) for $m$ even as
\begin{eqnarray}
&&\Delta V_{_S}(E)=\frac{E_x}\pi {\mathcal P}{\int_{-\infty }^\infty
}\frac{W_{_S}(E')} {(E'-E)(E'-E_{_F})}dE' =  \nonumber \\
&=&\frac{E_x}\pi {\mathcal P}{\int_0^\infty }\frac{ U^m\exp
  (-CU)}{(U^m+B^m)(U-E_{-}) (U+E_{_0})}dU+ \nonumber \\
&+&\frac{E_x}\pi {\mathcal P}{\int_0^\infty } \frac{ U^m\exp (-CU)}
{(U^m+B^m) (U+E_{+})(U+E_{_0})}dU
\label{integ}
\end{eqnarray}
and the same expression without exponential for the volume potential $W_{_V}(E)$,
given by equations (\ref{WV3}) and (\ref{symmetry}).\\
The integrand can be replaced by its expressions in terms of poles and
residues:
\begin{eqnarray}
&&\frac{E_x}\pi \frac{U^m}{(U^m+B^m)(U\mp E_{\mp
  })(U+E_{_0})}=\nonumber \\
&=&\frac 1
  \pi \left\{ \sum_{j=1}^m\frac{Res(p_j)}{U-p_j} + \frac {Res(\mp
  E_{_{\pm}})}{U\mp E_{\mp}} +\frac {Res(E_{_0})}{U+E_{_0}} \right\}
\label{subinteg}
\end{eqnarray}
where the $p_j$ are the $m$ zeroes of $(U^m+B^m)$ and the $Res(p_j)$ their
residue, that is :
\begin{eqnarray}
p_j &=&B\exp \left( i \frac{2j-1}{m}\cdot \pi \right) \\
Res(p_j) &=&\frac{E_x}{m}\frac{p_j}{(p_j\mp E_{\mp })(p_j+E_{_0})}
\end{eqnarray}
where $\pm E_{_{\mp}}$ and $-E_{_0}$ are the poles of $U\mp E_{_{\mp}}$
and $U+E_{_0}$, whereas $Res(\pm E_{_{\mp}})$ and $Res(-E_{_0})$ are their
residues respectively
\begin{eqnarray}
Res(\pm E_{_{\mp}})=\pm\frac{(E_{_{\mp}})^m}{(E_{_{\mp}})^m+B^m}\\
Res(-E_{_0})=\mp\frac{(E_{_0})^m}{(E_{_0})^m+B^m}
\end{eqnarray}
As was pointed out by Raynal \cite{raproc96} the contribution of each
complex pole $p_j$ to the surface dispersive integral is :
\begin{eqnarray}
&&\int_0^\infty\frac{Res(p_j)e^{-CU}}{U-p_j}dU=\nonumber \\
&=&Res(p_j)e^{-Cp_j}{\int_{-Cp_j}^\infty} \frac{\exp (-z)}zdz \equiv \nonumber\\
&\equiv&Res(p_j)e^{-Cp_j}E_1(-Cp_j)
\end{eqnarray}
where $E_1(z)$ is the {\em Exponential Integral Function} $E_1$
\cite{handbook}. For the real poles corresponding to the second term
in the right side of (\ref{subinteg}), the contribution of each one is given by:\\ 
\begin{eqnarray}
&&\int_0^\infty\frac{Res(\mp E_{_{\pm}})e^{-C U}}{U\mp
  E_{_{\mp}}}dU=\nonumber \\
&=&Res(\mp E_{_{\pm}})e^{\mp C E_{_{\mp}}}\mathcal{P}\int_{\mp C E_{_{\mp}}}^\infty
\frac {\exp (-x)}xdx \equiv \nonumber \\
&\equiv&-Res(\mp E_{_{\pm}})e^{\mp C E_{_{\mp}}} Ei(\pm C E_{_{\mp}})
\end{eqnarray}
where $Ei(x)$ is the {\em Exponential Integral Function} $Ei$
\cite{handbook}.\\
The contributions coming from the third term of the rigth side of eq.(\ref
{subinteg}) in the integral (\ref{integ}) cancel.\\
For the volume dispersive integral the exponential is missed in the
integrand, therefore each pole gives a divergent contribution.
Nevertheless their sum is a finite quantity, which can be calculated by
taking the proper limit.\\
We are quoting below exact expressions for the surface and volume
dispersive integrals for any even values of $m$ and $n$ in the surface and
volume imaginary potentials respectively. This make a difference to
ref. \cite{vaweto00}, where, in surface dispersive integral with $m=4$,
approximations were made limiting the energy range of validity for
given analytical expressions.\\
For the case of the dispersion relation using surface imaginary
potential $W_{_S}(E)$ with $m$ even according to equations (\ref{WS3}) and
(\ref{symmetry}), the dispersive correction is
\begin{eqnarray}
\Delta V_{_S}(E)&=&\frac{A}{\pi}\left\{\sum_{j=1}^{m}Z_j
~e^{-p_jC}E_1(-p_jC) \right. \nonumber \\
&-& Res(-E_+)~e^{CE_+}Ei(-CE_+) \nonumber \\
&-& \left. Res(E_-)~e^{-CE_-}Ei (CE_-) \Bigg\} \right.
\end{eqnarray}
where $Z_j$ comes from the sum of residues $Res(p_j)$ in the two
integrals (\ref{integ}) and is given by
\begin{equation}
Z_j=\frac{E_x}{m}\frac{p_j(2p_j+E_+-E_-)}{(p_j+E_{_0})(p_j+E_+)(p_j-E_-)}
\end{equation}
For the case of the dispersion relation using the volume imaginary
potential $W_{_V}(E)$ with n even according to equations (\ref{WV3}) and
(\ref{symmetry}), the dispersive correction is
\begin{eqnarray}
\Delta V_{_V}(E)&=&-\frac{A}{\pi}\left\{\sum_{j=1}^{n}Z_j
~ ln (-p_j)\right. \nonumber \\
&+& Res(-E_+)~ln{E_+}\nonumber \\
&+&\left. Res(E_-)~ln{|E_-|}\Bigg\} \right.
\end{eqnarray}

In conclusion, we have developed analytical solutions of dispersion
relations for the volume and surface terms of the OMP. The formulae
are compact and easy to implement in current generation of the optical
model parameter search codes. We stress that the solutions are valid
for any even exponent $n$ or $m$.\\

This work was supported by Junta de Andaluc\'{\i}a and  the Spanish
CICYT under Contracts PB1998-1111, FPA2001-0144-C05-03 and
FPA2001-4960-E and by the European Union under Contract FKIW-CT-2000-00107.

\bibliography{refer}

\begin{thebibliography}{11}
\expandafter\ifx\csname natexlab\endcsname\relax\def\natexlab#1{#1}\fi
\expandafter\ifx\csname bibnamefont\endcsname\relax
  \def\bibnamefont#1{#1}\fi
\expandafter\ifx\csname bibfnamefont\endcsname\relax
  \def\bibfnamefont#1{#1}\fi
\expandafter\ifx\csname citenamefont\endcsname\relax
  \def\citenamefont#1{#1}\fi
\expandafter\ifx\csname url\endcsname\relax
  \def\url#1{\texttt{#1}}\fi
\expandafter\ifx\csname urlprefix\endcsname\relax\def\urlprefix{URL }\fi
\providecommand{\bibinfo}[2]{#2}
\providecommand{\eprint}[2][]{\url{#2}}

\bibitem[{\citenamefont{Mahaux and Ng\^o}(1984)}]{mang84}
\bibinfo{author}{\bibfnamefont{C.}~\bibnamefont{Mahaux}} \bibnamefont{and}
  \bibinfo{author}{\bibfnamefont{H.}~\bibnamefont{Ng\^o}},
  \bibinfo{journal}{Nucl Phys. {\bf A431}, 486}  (\bibinfo{year}{1984}).

\bibitem[{\citenamefont{Mahaux and Sartor}(1987)}]{masa87np}
\bibinfo{author}{\bibfnamefont{C.}~\bibnamefont{Mahaux}} \bibnamefont{and}
  \bibinfo{author}{\bibfnamefont{R.}~\bibnamefont{Sartor}},
  \bibinfo{journal}{Nucl Phys. {\bf A468}, 193}  (\bibinfo{year}{1987}).

\bibitem[{\citenamefont{Johnson et~al.}(1987)\citenamefont{Johnson, Horen, and
  Mahaux}}]{johoma87}
\bibinfo{author}{\bibfnamefont{C.~H.} \bibnamefont{Johnson}},
  \bibinfo{author}{\bibfnamefont{D.~J.} \bibnamefont{Horen}}, \bibnamefont{and}
  \bibinfo{author}{\bibfnamefont{C.}~\bibnamefont{Mahaux}},
  \bibinfo{journal}{Phys. Rev. C {\bf 36}, 2252}  (\bibinfo{year}{1987}).

\bibitem[{\citenamefont{Mahaux and Sartor}(1991{\natexlab{a}})}]{masa91}
\bibinfo{author}{\bibfnamefont{C.}~\bibnamefont{Mahaux}} \bibnamefont{and}
  \bibinfo{author}{\bibfnamefont{R.}~\bibnamefont{Sartor}},
  \bibinfo{journal}{Nucl. Phys. {\bf A528}, 253}
  (\bibinfo{year}{1991}{\natexlab{a}}).

\bibitem[{\citenamefont{Mahaux and Sartor}(1991{\natexlab{b}})}]{masa91rev}
\bibinfo{author}{\bibfnamefont{C.}~\bibnamefont{Mahaux}} \bibnamefont{and}
  \bibinfo{author}{\bibfnamefont{R.}~\bibnamefont{Sartor}},
  \emph{\bibinfo{title}{Advances in Nuclear Physics}},
  vol.~\bibinfo{volume}{20} (\bibinfo{publisher}{(edited by {J.W.~Negele} and
  {E.~Vogt}. Plenum, New York)}, \bibinfo{year}{1991}{\natexlab{b}}).

\bibitem[{\citenamefont{Delaroche et~al.}(1989)\citenamefont{Delaroche, Wang,
  and Rapaport}}]{dewara89}
\bibinfo{author}{\bibfnamefont{J.~P.} \bibnamefont{Delaroche}},
  \bibinfo{author}{\bibfnamefont{Y.}~\bibnamefont{Wang}}, \bibnamefont{and}
  \bibinfo{author}{\bibfnamefont{J.}~\bibnamefont{Rapaport}},
  \bibinfo{journal}{Phys. Rev. C {\bf 39}, 391}  (\bibinfo{year}{1989}).

\bibitem[{\citenamefont{Brown and Rho}(1981)}]{brrh81}
\bibinfo{author}{\bibfnamefont{G.~E.} \bibnamefont{Brown}} \bibnamefont{and}
  \bibinfo{author}{\bibfnamefont{M.}~\bibnamefont{Rho}}, \bibinfo{journal}{Nucl
  Phys. {\bf A372}, 397}  (\bibinfo{year}{1981}).

\bibitem[{\citenamefont{VanderKam et~al.}(2000)\citenamefont{VanderKam, Weisel,
  and Tornow}}]{vaweto00}
\bibinfo{author}{\bibfnamefont{J.~M.} \bibnamefont{VanderKam}},
  \bibinfo{author}{\bibfnamefont{G.~J.} \bibnamefont{Weisel}},
  \bibnamefont{and} \bibinfo{author}{\bibfnamefont{W.}~\bibnamefont{Tornow}},
  \bibinfo{journal}{J. Phys. G:Nucl.Part.Phys. {\bf 26}, 1787}
  (\bibinfo{year}{2000}).

\bibitem[{\citenamefont{Capote et~al.}(2001)\citenamefont{Capote, Molina, and
  Quesada}}]{camoqu01}
\bibinfo{author}{\bibfnamefont{R.}~\bibnamefont{Capote}},
  \bibinfo{author}{\bibfnamefont{A.}~\bibnamefont{Molina}}, \bibnamefont{and}
  \bibinfo{author}{\bibfnamefont{J.~M.} \bibnamefont{Quesada}},
  \bibinfo{journal}{J. Phys. G:Nucl.Part.Phys. {\bf 27}, B15}
  (\bibinfo{year}{2001}).

\bibitem[{\citenamefont{Raynal}(1997)}]{raproc96}
\bibinfo{author}{\bibfnamefont{J.}~\bibnamefont{Raynal}}, in
  \emph{\bibinfo{booktitle}{Proceedings of the Specialists' Meeting on the
  Nucleon Nucleus Optical Model up to 200 MeV, Bruy\`eres-le-Ch\^atel, 1996.
  Available at http://www.nea.fr/html/science/om200/.}} (\bibinfo{year}{1997}).

\bibitem[{\citenamefont{Abramowitz and Stegun}(1968)}]{handbook}
\bibinfo{author}{\bibfnamefont{M.}~\bibnamefont{Abramowitz}} \bibnamefont{and}
  \bibinfo{author}{\bibfnamefont{I.}~\bibnamefont{Stegun}},
  \emph{\bibinfo{title}{{\it Handbook of Mathematical Functions, Applied
  Mathematics Series, vol. 55}}} (\bibinfo{publisher}{(Washington: National
  Bureau of Standards; reprinted by Dover Publications, New York)},
  \bibinfo{year}{1968}).

\end{thebibliography}

\end{document}